\documentclass[11pt]{article}
\usepackage{float,amssymb}
\usepackage{graphicx}
\newcommand{\beq}{\begin{equation}}
\newcommand{\eeq}{\end{equation}}
\newcommand{\beqa}{\begin{eqnarray}}
\newcommand{\eeqa}{\end{eqnarray}}

\newcommand{\betae}{\beta_{\varepsilon}}

\newcommand{\diffesq}{{\cal E}^2 - m_e^2}
\newcommand{\diffeksq}{({\cal E} - k^{\rm max})^2 - m_e^2}
\newcommand{\diffo}{E_e^{\rm max} - E_e}
\newcommand{\diffod}{E_e^{\rm max} - E_e - \Delta E}
\newcommand{\difftwd}{(E_e^{\rm max} - E_e)^2 - (\Delta E)^2}
\newcommand{\diffthd}{(E_e^{\rm max} - E_e)^3 - (\Delta E)^3}
\newcommand{\difffod}{(E_e^{\rm max} - E_e)^4 - (\Delta E)^4}

\newcommand{\vs}{\vspace{-0.20cm}}
\begin{document}


\begin{flushright}
{\tiny  UK/TP-2004-15} \\
{\tiny  HISKP-TH-04/12} \\
\end{flushright}

\vspace{.6in}

\begin{center}

\bigskip

{{\Large\bf Radiative tritium  \boldmath{$\beta$}--decay 
and the neutrino mass}}

\end{center}

\vspace{.3in}

\begin{center}
{{\large 
Susan Gardner$^\dagger$\footnote{email: gardner@pa.uky.edu},
V\'eronique Bernard$^\star$\footnote{email: bernard@lpt6.u-strasbg.fr},
Ulf-G. 
Mei{\ss}ner$^\ddagger$$^\ast$\footnote{email: meissner@itkp.uni-bonn.de}
}}

\vspace{1cm}

$^\dagger${\it Department of Physics and Astronomy, University of Kentucky\\
Lexington, Kentucky 40506-0055, USA}\\

\bigskip

$^\star${\it Universit\'e Louis Pasteur, Laboratoire de Physique
            Th\'eorique\\ 3-5, rue de l'Universit\'e,
            F--67084 Strasbourg, France}

\bigskip

$^\ddagger${\it Universit\"at Bonn,
Helmholtz--Institut f\"ur Strahlen-- und Kernphysik (Theorie)\\
Nu{\ss}allee 14-16,
D-53115 Bonn, Germany}

\bigskip

$^\ast${\it Forschungszentrum J\"ulich, Institut f\"ur Kernphysik 
(Theorie)\\ D-52425 J\"ulich, Germany}

\bigskip

\bigskip

\end{center}

\vspace{.4in}

\thispagestyle{empty} 

\begin{abstract}\noindent 
The shape of the electron energy spectrum in $^3$H $\beta$-decay
permits a direct assay of the absolute scale of the neutrino mass; a 
highly accurate theoretical description of the electron energy spectrum
is necessary to the empirical task. 
We update Sirlin's calculation of the outer radiative correction
to nuclear $\beta$-decay to take into account the non-zero energy resolution
of the electron detector. In previous $^3$H $\beta$-decay studies the
outer radiative corrections were neglected all together; only Coulomb
corrections to the spectrum were included. 
This neglect artificially pushes $m_\nu^2 < 0$ in a 
potentially significant
way. We present a computation of the theoretical spectrum appropriate
to the extraction of the neutrino mass in the sub-eV regime. 
\end{abstract}

\vfill

\pagebreak

Empirical evidence of neutrino oscillations in 
atmospheric, solar, and reactor 
neutrino data~\cite{superk,sno,kamland} compels the existence of 
non-zero neutrino masses, yet such experiments are insensitive
to the absolute scale of a neutrino mass, for the oscillation
experiments determine $\Delta m_{ij}^2\equiv m_i^2 - m_j^2$, 
where $m_i$ is the mass of neutrino $i$. To determine 
the absolute value of the neutrino mass 
requires different methods. 
Cosmological constraints on the neutrino mass do exist~\cite{pierce},
though our focus shall be on the study of the 
electron energy spectrum  in tritium $\beta$-decay
near its endpoint, as this represents the most
sensitive terrestrial measurement. The spectrum shape constrains 
the mass of the neutrino, be it of Dirac or Majorana character, and
the inferred mass is insensitive to phases in the neutrino
mixing matrix --- in contradistinction to the constraint on
the neutrino mass from neutrinoless double $\beta$-decay. 
An accurate theoretical description of the expected electron
energy spectrum is crucial to the determination of the neutrino mass; 
this demand grows as the sensitivity of the experiments increase.
Indeed, future studies expect to probe the neutrino mass at the sub-eV 
level~\cite{katrin}. It is our purpose to realize a theoretical
form of the requisite accuracy, though we shall begin by 
describing the form used in earlier tritium experiments.

With an anti-electron neutrino of mass $m_\nu$, neglecting neutrino 
mixing for simplicity, the
Fermi form of the 
electron energy spectrum for tritium $\beta$-decay is~\cite{robertson}
\begin{equation}
\frac{d\Gamma_{F}}{dE_e} = \frac{G_F^2}{2\pi^3} |{\cal M}|^2 
F(Z,R_{\rm e},E_e) 
p_e E_e (E_e^{\rm max} - E_e)\sqrt{(E_e^{\rm max} - E_e)^2 - m_\nu^2} \,,
\label{fermiform}
\end{equation}
where $G_F$ is the Fermi constant, $p_e$, $E_e$, and $E_e^{\rm max}$ are
the momentum, energy, and maximum endpoint energy, respectively, 
of the electron, and $|{\cal M}|^2$ is the absolute
square of the nuclear matrix element, with 
$|{\cal M}|^2\sim 5.3$. 
A form of this ilk has been used to bound $m_\nu$ in previous 
experimental analyses of molecular tritium 
$\beta$-decay~\cite{robprl,exp1,exp2,exp3,wein94,exp4,troitsk,weinplb}. 
Following the usual practice, we include a non-zero neutrino mass 
in the phase space contribution only. 
We set $\hbar=c=1$ throughout. 
The Fermi function, $F(Z,R_{\rm e},E_e)$, 
captures the correction
due to the Coulomb interactions of the electron with the charge $Ze$
of the daughter nucleus~\cite{coulomb}. We adopt the usual 
expression~\cite{atdata}, derived from the solutions of the Dirac
equation for the point-nucleus potential $-Z\alpha/r$ evaluated
at the nuclear radius $R_{\rm e}$~\cite{texts}; 
it differs from unity by a contribution of ${\cal O}(\alpha)$. 
The Fermi function includes  
the dominant electromagnetic effect, though  
an accurate extraction, or bound, of the 
neutrino mass does demand the inclusion of the remaining
${\cal O}(\alpha)$ correction. 
We shall demonstrate this point explicitly. 
This last effect, termed the radiative correction,
is conventionally separated into an ``inner'' piece $\Delta_R$,
which is absorbed in $|{\cal M}|^2$, as it is energy independent 
and thus of no consequence
to our current study, and an ``outer'' 
piece $\delta_R$~\cite{sirlin}. The outer radiative correction
applied in $\beta$-decay studies is due to Sirlin~\cite{sirlin}: 
\begin{equation}
\frac{d\Gamma}{dE_e} = \frac{d\Gamma_0}{dE_e}
\left(1 
+ \frac{\alpha}{2\pi} g_S(E_e,E_e^{\rm max})\right)\,, 
\end{equation} 
where 
$d\Gamma_0/dE_e = G_F^2|{\cal M}|^2 p_e E_e (E_e^{\rm max} - E_e)^2/(2\pi^3)$, 
and, noting $\beta=p_e/E_e$,  
\begin{eqnarray}
g_S(E_e,E_e^{\rm max})&=& 3 \ln\left(\frac{M}{m_e}\right) 
- \frac{3}{4} 
+ 4 \left[\frac{\tanh^{-1} \beta}{\beta} - 1 \right]\left[
\frac{(E_e^{\rm max} - E_e)}{3E_e} - \frac{3}{2} + 
\ln\Big\{\frac{2(E_e^{\rm max} - E_e)}{m_e}\Big\}\right]  \nonumber \\
&& + \frac{4}{\beta}
L\left(\frac{2\beta}{1+\beta}\right) +
\frac{\tanh^{-1} \beta}{\beta}\left[2(1+\beta^2) + 
\frac{(E_e^{\rm max} - E_e)^2}{6E_e^2} - 4 \tanh^{-1} \beta\right] \,,
\label{sirlin}
\end{eqnarray} 
with 
$M$ and $m_e$ the proton and electron mass, respectively, 
and $L(x)$ the Spence function. As in 
Ref.~\cite{sirlin}, we neglect terms of relative order
$\alpha (E_e/M)\ln(M/E_e)$, $\alpha E_e/M$, and smaller, throughout. 
The total ${\cal O}(\alpha)$ correction is given by 
$F(Z,R_{\rm e},E_e) - 1 + (\alpha/2\pi)g_S(E_e,E_e^{\rm max})$. 
Note that $\delta_R$ results from {\it averaging}
$(\alpha/2\pi)g_S(E_e,E_e^{\rm max})$ over the
electron energy spectrum. To assess the relative sizes of $F$
and $g_S$, we note that in the endpoint region of tritium $\beta$-decay, for 
$E_e - m_e = 18.5$ keV, e.g., $F \sim 1.19$, whereas 
$(\alpha/2\pi)g_S \sim 0.02$. 

\begin{figure}[ht]
\begin{center}
\includegraphics[width=4.6in]{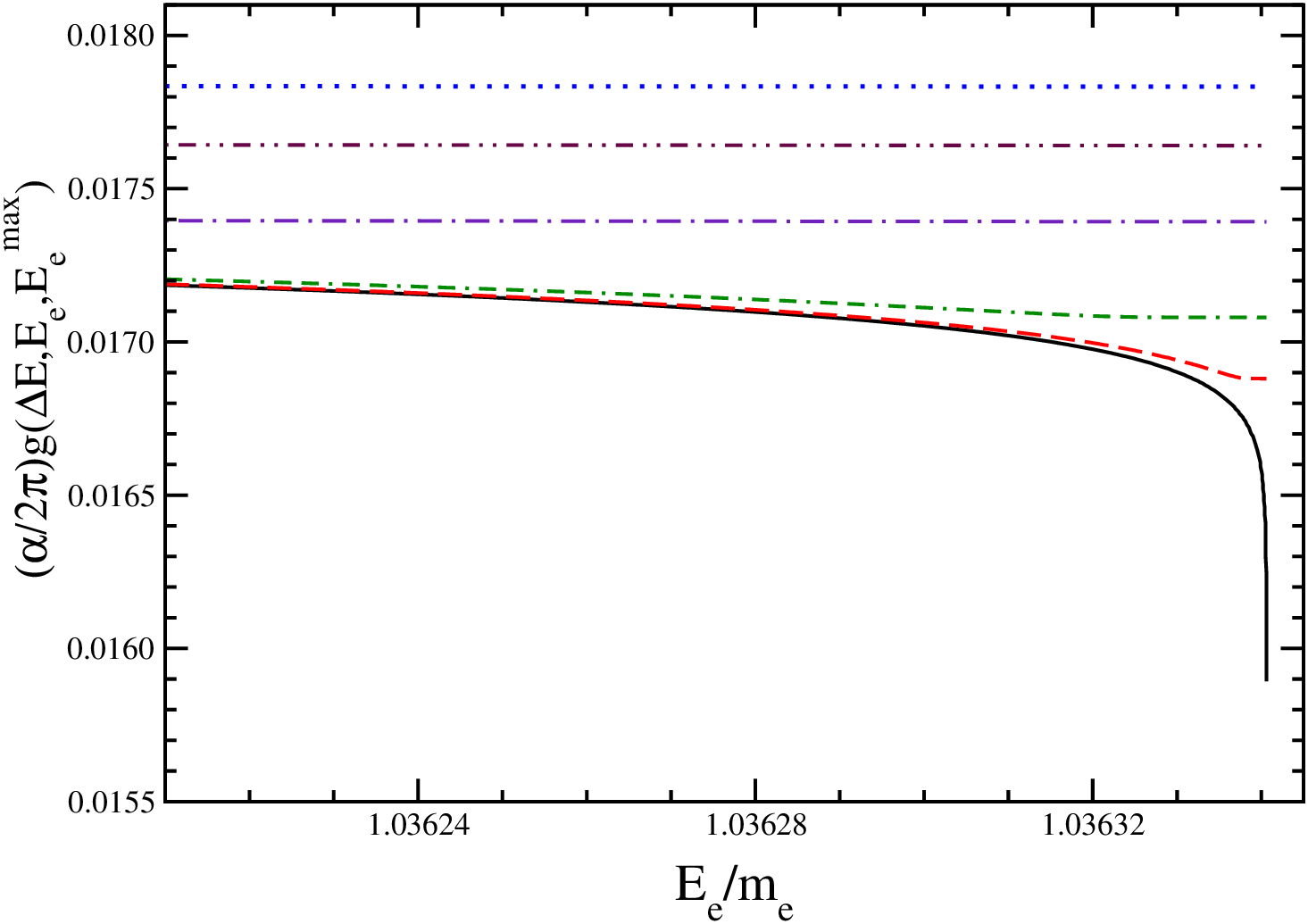}
\vspace{-0.2cm}
\caption{The ``outer'' radiative correction
$g(\Delta E,E_e,E_e^{\rm max})$ as a function of the electron
detector resolution $\Delta E$ and the electron
energy $E_e$ in the endpoint region of tritium $\beta$-decay. 
The solid line is Sirlin's result~\cite{sirlin}, for which 
$\Delta E=0$. The dotted line corresponds 
to Vogel's result~\cite{vogel}, for which 
$\Delta E=E_e^{\rm max} - m_e$. 
The remaining curves correspond to 
$\Delta E=1$~\cite{katrin}, $6$~\cite{wein94}, $100$, 
and $1000$ eV, respectively,  
moving in sequence from the solid line to the dotted one.  
The chosen values of $\Delta E$ correspond to those of the planned
and recent experiments to which we refer.
\label{fig:geff}}
\end{center}
\end{figure}

Since the absolute neutrino mass scale is inferred through the
shape of the electron energy spectrum in the endpoint region, it
is crucial to predict the shape of the theoretical spectrum with 
high accuracy. To this end, we update the calculation of Ref.~\cite{sirlin}
to take the energy resolution of the electron detector into account. 
To understand the significance of this, we recall that Sirlin's function
contains not only virtual photon corrections to the $\beta$-decay
process but also bremsstrahlung contributions, 
to yield an additional real photon in the final state. 
Only their sum is infrared finite; 
the infrared divergence
in each contribution is regulated by giving the photon a small
mass $\lambda$, with the $\lambda\to 0$ limit to be taken after the sum
has been computed and the infrared divergent pieces cancelled. 
The finite portion of the bremsstrahlung contribution
is sensitive to the precise manner
in which the experiment is effected. 
In Sirlin's function the energy resolution of the electron detector
is implicitly assumed to be zero; that is, the $e^-$ and $\gamma$
are always distinguishable. A consequence of this is that 
Eq.~(\ref{sirlin}) contains a logarithmic 
divergence as $E_e \to E_e^{\rm max}$. 
This singularity can
be removed by including soft photon contributions to all
orders in perturbation theory~\cite{wurepko}; however, it can
also be removed by taking the energy resolution of the
detector into account, as we consider here. 
The bremsstrahlung 
contribution comes from integrating the photon energy 
over its entire kinematic range, namely 
\begin{equation}
\frac{\alpha}{2\pi} \left( \frac{d\Gamma_0}{dE_e} \right)
g_{S,b}(E_e,E_e^{\rm max}) = 
\int_\lambda^{E_e^{\rm max} - E_e} dE_\gamma 
\left(\frac{d^2 \Gamma_\gamma(E_e,E_\gamma)}{d E_e d E_\gamma}\right)
\,,
\end{equation}
where 
$d^2 \Gamma_\gamma(E_e,E_\gamma)/d E_e d E_\gamma$ is 
the doubly differential decay rate for the radiative $\beta$-decay
process. 
Nevertheless, the detector energy resolution $\Delta E$ can, in principle, 
influence 
the shape of the electron energy spectrum. That is, for 
$E_\gamma \le \Delta E$, the electron and photon {\it cannot} be
distinguished; indeed, this is precisely why 
the bremsstrahlung contribution can enter to render an infrared-finite, 
${\cal O}(\alpha)$
radiative correction to $\beta$-decay. In this event 
the detector records the sum of the electron and photon energies
as the ``electron'' energy. 
For $E_\gamma > \Delta E$, however, 
the electron and photon energies are distinguishable, and their
energies can be recorded separately. To separate the contributions
we note that 
the {\it total} bremsstrahlung contribution to $\delta_R$ 
must be insensitive to such
experimental details. In passing, we note related discussions of the 
impact of the detection threshold for bremsstrahlung photons in the
radiative corrections to $\nu$ capture on 
deuterium~\cite{vogel,towner,parke,kurylov}, as well as to $\nu-e$ 
scattering~\cite{passera}. 
The bremsstrahlung contribution
to the total outer radiative correction $\delta_R$ --- we retain
the photon mass $\lambda$ throughout --- 
is
\begin{equation}
\tilde \delta_{R,b}
= \int_{m_e}^{E_e^{\rm max} - \lambda} dE_e 
\int_\lambda^{E_e^{\rm max} -  E_e} dE_\gamma \,f(E_e,E_\gamma) \,, 
\label{sirlinb}
\end{equation}
where $f(E_e,E_\gamma)\equiv d^2 \Gamma_\gamma(E_e,E_\gamma)/dE_e dE_\gamma$. 
To find $\delta_R$ we must divide $\tilde \delta_{R,b}$
by a normalization factor $N$, determined by integrating Eq.~(\ref{fermiform})
over the allowed phase space with $F=1$ and $m_\nu=0$. 
However, if the detector energy resolution is ``infinite,'' that is, if
the electron detector always records ${\cal E}=E_e + E_\gamma$~\cite{vogel}, 
the total bremstrahlung contribution can be rewritten as 
\begin{equation}
\tilde \delta_{R,b} = 
\int_{m_e+\lambda}^{E_e^{\rm max}} d{\cal E}
\int_\lambda^{{\cal E} - m_e} dE_\gamma 
\,f({\cal E} - E_\gamma,E_\gamma) \,.
\label{vogelb}
\end{equation}
In both Eqs.~(\ref{sirlinb}) and (\ref{vogelb}), 
the integration over $E_\gamma$ yields the 
bremstrahlung correction to the electron energy spectrum. Although 
$\tilde\delta_{R,b}$ is universal, the shape correction to the electron 
energy spectrum is
not. Let us now determine the shape correction for a finite energy
resolution $\Delta E$. 
That is, for $E_\gamma \le \Delta E$, ${\cal E} = E_e + E_\gamma$
is recorded, whereas for $E_\gamma > \Delta E$, $E_e$ is recorded. 
We can reorganize the total bremsstrahlung contribution in the following way:
\begin{eqnarray}
\tilde \delta_{R,b} &=& 
\int_{m_e+\lambda}^{\Delta E + m_e} d{\cal E}
\int_\lambda^{{\cal E} - m_e} dE_\gamma \,f({\cal E} - E_\gamma,E_\gamma)
+ 
\int_{\Delta E + m_e}^{E_e^{\rm max}} d{\cal E}
\int_\lambda^{\Delta E} dE_\gamma 
\,f({\cal E} - E_\gamma,E_\gamma) \nonumber \\
&& + 
\int_{m_e}^{E_e^{\rm max} - \Delta E} dE_e 
\int_{\Delta E}^{E_e^{\rm max} - E_e} dE_\gamma \,f(E_e,E_\gamma) \,.
\label{deltaE}
\end{eqnarray}
Note that letting $\Delta E\to \lambda$ yields Eq.~(\ref{sirlinb}),
whereas letting $\Delta E\to E_e^{\rm max} - m_e$ yields Eq.~(\ref{vogelb}).
If $\Delta E$ has a non-infinitesimal value, then infrared
divergences are restricted to the first two terms --- in the third,
we may let $E_\gamma = \sqrt{k^2 + \lambda^2}\to k$ with impunity. 
Thus we introduce 
\begin{eqnarray}
&& \frac{\alpha}{2\pi}\frac{d\Gamma_0}{d{\cal E}} 
{\cal I}_\lambda(k^{\rm max},{\cal E}) = 
\int_\lambda^{k^{\rm max}} dE_\gamma \,
f({\cal E} - E_\gamma, E_\gamma) \,\nonumber\\
\vspace{-0.2cm}
&&\!\!\!\!\!\!\!\!\!\!\!\!\!\!\!\!\!\!\!\!\!\!\!\!\!\!\!\!\!\!\!\!\!\!\!\!\!\!
\!\!\!\!\!\!\!
\hbox{and} \nonumber \\
\vspace{-0.2cm}
&& \frac{\alpha}{2\pi}\frac{d\Gamma_0}{dE_e} {\cal I}(\Delta E, E_e) =
\int_{\Delta E}^{E_e^{\rm max} -E_e} dE_\gamma \,
f(E_e, E_\gamma) \to 
\int_{\Delta E}^{E_e^{\rm max} -E_e} dk \, f(E_e, k) 
\label{ints}
\end{eqnarray}
to realize
\begin{eqnarray}
g_b(\Delta E,E_e,E_e^{\rm max}) 
&=&
\Theta(\Delta E + m_e - E_e) {\cal I}_{\lambda}(k^{\rm max} = E_e - m_e, E_e)
\nonumber \\
&+&
\Theta(E_e - (\Delta E + m_e) ) 
{\cal I}_{\lambda}(k^{\rm max} = \Delta E, E_e) \nonumber \\
&+&
\Theta(E_e^{\rm max} - \Delta E - E_e) ) {\cal I}(\Delta E, E_e)\,.
\label{gdeltaE}
\end{eqnarray}
With this, we determine that 
the radiative correction to the electron energy spectrum is 
\begin{equation}
g(\Delta E,E_e,E_e^{\rm max}) =  
g_b(\Delta E,E_e,E_e^{\rm max}) + g_v(E_e) \,,
\label{gfinal}
\end{equation}
where the virtual photon contribution $g_v\equiv 2{\cal A}-3/4$, 
with ${\cal A}$ as reported in Eq.~(16) of Ref.~\cite{kurylov}. 
To compute the integrals of Eq.~(\ref{ints}) and thus 
$g_b(\Delta E,E_e,E_e^{\rm max})$ we adapt the computation of
radiative neutron $\beta$-decay in Ref.~\cite{bgmz} to this case. 
In specific, we use the absolute squared matrix elements of Eqs.~(13-16),
and Eq.~(20), dividing the latter by $8 M^2$, in that work to determine 
$f(E_e,E_\gamma)$. In doing the integrals, we can neglect the 
recoil corrections to the phase space, so that 
$E_\nu = E_e^{\rm max} - E_e - k$. As a result,  
the dependence on the weak, hadron coupling constants
is captured by $g_V^2 + 3 g_A^2$, which we replace by $|{\cal M}|^2$; 
we also update 
the mass factors which appear as appropriate. 
We have verified that our 
computation of $g_b(\Delta E=0, E_e, E_e^{\rm max})$ using this 
procedure and 
Eq.~(\ref{sirlinb}), in concert
with $g_v$ of Ref.~\cite{kurylov}, yields Sirlin's result~\cite{sirlin}, 
Eq.~(\ref{sirlin}). For finite $\Delta E$ we find the following
results: 
\begin{eqnarray}
&&{\cal I}_\lambda(k^{\rm max},{\cal E}) =
6 
- 4\left(1 - \frac{\tanh^{-1} \betae}{\betae}\right)
\ln \left(\frac{2k^{\rm max}}{\lambda}\right) 
+ \frac{2 \tanh^{-1} \betae}{\betae}
+ \frac{2}{\betae} L\left(\frac{2\betae}{1+ \betae}\right) \nonumber\\
&& - \frac{2}{\betae} (\tanh^{-1} \betae)^2 
+ \frac{2}{\betae} I_k^{-1} 
- \frac{2}{{\cal E}\betae} I_k^{0} 
+ \frac{1}{{\cal E}^2\betae} I_k^{1} 
- 4 \frac{\sqrt{\diffeksq}}{\sqrt{\diffesq}}  
\nonumber \\
&&
+ 4 \frac{1}{\betae} \ln\left(
\frac{{\cal E} - k^{\rm max} - \sqrt{\diffeksq}}{{\cal E} - \sqrt{\diffesq}}
\right) \nonumber \\
&& + 4 \ln \left(
\frac{ {\cal E}^2 - m_e^2 - {\cal E}k^{\rm max} 
+ \sqrt{\diffesq}\cdot\sqrt{\diffeksq}}{2 ({\cal E}^2 - m_e^2)}
\label{res1}
\right)
\,,
\end{eqnarray}
noting $\betae =\sqrt{{\cal E}^2 - m_e^2}/{\cal E}$, and 
\begin{equation}
I_k^{-1} = \int_0^{k^{\rm max}} dk\, \frac{1}{k} 
\ln\left(\frac{1+\beta_k}{1+\betae}\cdot \frac{1-\betae}{1-\beta_k}\right) 
\;\;\;,\;\;\;
I_k^n = \int_0^{k^{\rm max}} dk\, k^n 
\ln\left(\frac{1+\beta_k}{1-\beta_k}\right) 
\;\;\;\hbox{with}\;\;\; n= 0,1\,,
\label{res2}
\end{equation}
where  $\beta_k=\sqrt{({\cal E} - k)^2 - m_e^2}/({\cal E} -k)$.
(We note that $I_k^n$ can be brought to closed form via the
substitution $t=\beta_k$, though we omit the resulting
expressions here.) Moreover, 
\begin{eqnarray}
&&{\cal I}(\Delta E, E_e)= -4 \ln\left(\frac{\diffo}{\Delta E}\right)
- 4 \frac{(\diffod)}{E_e} + 8 \frac{(\diffod)}{(\diffo)} \nonumber\\
&&+ 4 \frac{(\difftwd)}{E_e(\diffo)} 
 - 2 \frac{(\difftwd)}{(\diffo)^2} 
-\frac{4}{3}\frac{(\diffthd)}{E_e(\diffo)^2} \nonumber \\
&&+ \frac{2}{\beta} \tanh^{-1}\beta \Bigg[
2 \ln\left(\frac{\diffo}{\Delta E}\right)
+2 \frac{(\diffod)}{E_e} +  \frac{(\difftwd)}{2E_e^2} \nonumber \\
&&-4 \frac{(\diffod)}{(\diffo)} 
-2 \frac{(\difftwd)}{E_e (\diffo)}
-\frac{2}{3} \frac{(\diffthd)}{E_e^2 (\diffo)} \nonumber \\
&& + \frac{(\difftwd)}{(\diffo)^2}  
+ \frac{2}{3} \frac{(\diffthd)}{E_e(\diffo)^2} 
+ \frac{1}{4} \frac{(\difffod)}{E_e^2(\diffo)^2} 
\Bigg] \,.
\label{res3}
\end{eqnarray}
We note that the $\ln \lambda$ term in 
${\cal I}_\lambda(k^{\rm max},{\cal E})$ 
cancels the 
concommitant infrared divergent term in $g_v$ to yield a finite
result in the $\lambda\to 0$ limit, irrespective of the detector 
energy resolution $\Delta E$. 
Using these formulae, we find our 
$g(\Delta E=E_e^{\rm max}-m_e, E_e, E_e^{\rm max})$ is in
accord with the result of Vogel, Ref~\cite{vogel}. 
We report $g(\Delta E,E_e,E_e^{\rm max})$ in the endpoint
region of tritium $\beta$-decay, which results from 
Eqs.~(\ref{ints}-\ref{res3}), 
in Fig.~\ref{fig:geff}.
We have verified that the
integration of $(d\Gamma_0/dE_e)g(\Delta E,E_e,E_e^{\rm max})$ over
$E_e$ yields a universal value of $\tilde \delta_{R,b}$ for all
$\Delta E$.  
The inclusion of a finite value of $\Delta E$ removes the logarithmic
divergence in Sirlin's function as $E_e \to E_e^{\rm max}$. The 
numerical shifts associated with the inclusion of the $\Delta E$ 
dependence generally are crudely comparable
in size to the leading ${\cal O}(Z\alpha^2)$ correction~\cite{sirzuc},
$-Z\alpha^2 \ln(M/m_e)$, though the latter, of course, contains
no $E_e$ dependence.
\vspace{1.1cm}

\vspace{-1.0cm}
We can now proceed to evaluate the changes these theoretical
corrections make to the shape assumed in previous
experimental assays of the $\nu$ mass. 
For definiteness, we also include the leading-order recoil
corrections to the electron energy spectrum. We may once again adapt
the results from neutron decay to this case. 
We adopt 
the notation of Bender et al. in Ref.~\cite{recoil}, though
the couplings of the hadronic weak current are now 
nuclear form factors evaluated
at zero momentum transfer.  
Noting Ref.~\cite{recoil}, 
we replace the absolute, squared nuclear transition matrix element,
which we have taken to be 
$|{\cal M}|^2 = |{\cal M}_0|^2 = g_V^2 + 3 g_A^2$, 
with 
\begin{equation}
|{\cal M}|^2 \to |{\cal M}_0|^2 (1 + {\cal R}) \;, 
\label{recsub}
\end{equation}
where 
\begin{eqnarray}
&& \!\!\!\!\!\!\!\! 
{\cal R} = \frac{1}{( g_{V}^2 + 3 g_{A}^2 )}\Bigg[
g_{A} f_{2} 
\left( -4 \frac{m_e^2}{M_A E_e} - 4\frac{E_e^{\rm max}}{M_A} + 
8\frac{E_e}{M_A}\right) + g_{V}^2 \left( 2 \frac{E_e}{M_A} \right)
\nonumber \\
&& 
\!\!\!\!\!\!\!\! + g_{A}^2 
\left( -2 \frac{m_e^2}{M_A E_e} - 
2\frac{E_e^{\rm max}}{M_A} + 10\frac{E_e}{M_A}\right)
+ g_{V} g_{A}  
\left( -2 \frac{m_e^2}{M_A E_e} - 2\frac{E_e^{\rm max}}{M_A} + 
4\frac{E_e}{M_A}\right)  \Bigg] 
\,.
\label{recoil}
\end{eqnarray}
Note that $M_A =2809.4319$ MeV~\cite{3hdata} is the tritium mass and 
$E_e^{\rm max} - m_e = 18.57$ keV~\cite{weinplb}, so that 
$E_e^{\rm max}/M_A \sim 1.9 \cdot 10^{-4}$. 
Since $|{\cal M}_0|^2$ can be absorbed into the overall normalization
of the decay rate, the function ${\cal R}$ 
represents the first appearance of nuclear-structure effects
in the prediction of the electron-energy spectrum in tritium $\beta$-decay. 
The form factors which enter
are largely determined by the symmetries of
the Standard Model (SM), so that the subsequent uncertainty in the predicted
recoil correction, which is itself of small numerical size, is very small. 
In writing Eq.~(\ref{recoil}), we have assumed the validity of the 
conserved-vector-current (CVC) 
hypothesis and have neglected the form factors associated 
with second-class-current contributions. In the context of the SM,
this is tantamount to neglecting the effects of isospin violation,
so that the recoil term is subject to corrections of ${\cal O}(1\%)$. 
The vector coupling $g_V$ is also unity
by the CVC hypothesis; the computed correction due to charge-symmetry
breaking in the overlap of the $^3$H-$^3$He wave functions, due to
Towner, is $\delta_c =0.06\%$~\cite{simpson}; we note
$g_V^2 = {g_V^\prime}^2(1 - \delta_c)$, where ${g_V^\prime}^2$
absorbs the inner radiative correction $\Delta_R$ and $|V_{ud}|^2$,
with $V_{ud}$ a CKM matrix element.  
The CVC hypothesis 
determines the weak-magnetism coupling $f_2$ from the 
measured $^3$H and $^3$He magnetic moments~\cite{3hdata}, to yield 
$f_2/g_V^\prime= -3.0533$; we ignore the possibility of a 
inner radiative correction idiosyncratic to $f_2$. 
The $^3$H half-life determines 
$g_{V}^2 + 3 g_{A}^2$  up to corrections of recoil order; this
is sufficient to determine the couplings which appear in the
recoil-order expression. In specific, we have~\cite{simpson}
$(1 - \delta_c + 3 g_A^2/{g_V^\prime}^2)^{-1} = 
G_F^2{g_V^\prime}^2 f(1 + \delta_R) t_{1/2} /K$ with 
${g_V^\prime}^2 = (1 + \Delta_R)|V_{ud}|^2$ and 
$K=2\pi^3 \ln 2/m_e^5$. 
We use $G_F$, $\alpha$, $m_e$, and $\hbar$
as given in Ref.~\cite{pdg2002}, $\Delta_R=0.0240$, $V_{ud}=0.9740$  
as in Ref.~\cite{wein98}, and the half-life 
$t_{1/2}=12.3$ yrs as recommended in Ref.~\cite{3hdata}. Finally
we use the integral of $F+(\alpha/2\pi)g_S$, noting Eq.~(4) 
of Ref.~\cite{atdata} for $F$, over the allowed phase space to fix
$f(1 + \delta_R)$, for which we find $2.9109 \cdot 10^{-6}$.
We use $R_e=1.68$ fm throughout~\cite{collard,3hdata}. 
This yields $g_A/g_V^\prime=1.22$; note that this ratio of
couplings implicitly contains the quenching of the Gamow-Teller matrix
element due to nuclear structure effects. 
\begin{figure}[ht]
\begin{center}
\includegraphics[width=4.6in]{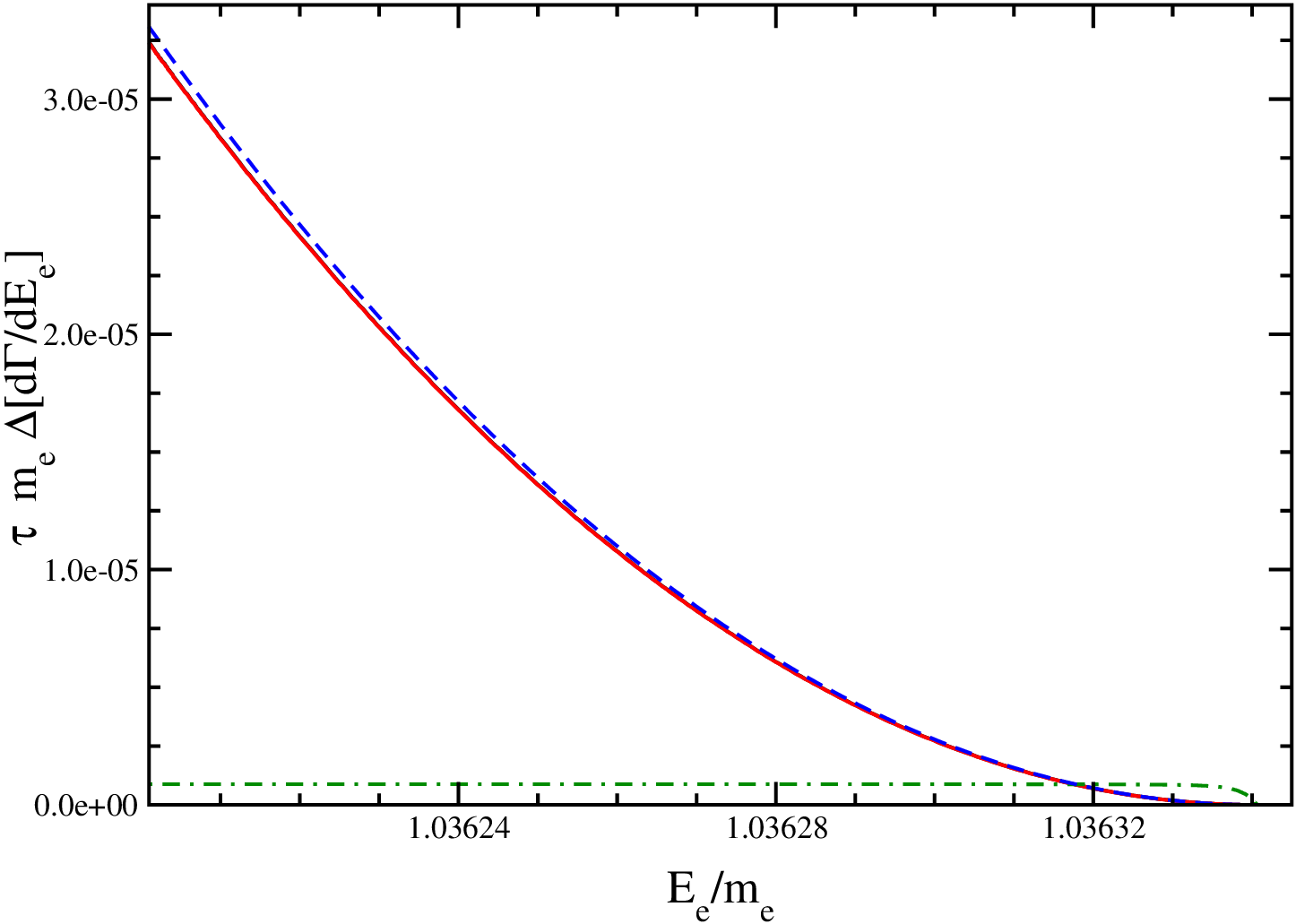}
\vspace{-0.2cm}
\caption{The change in the electron energy spectrum, 
$d\Gamma/dE_e$, upon the inclusion of $g(\Delta E,E_e,E_e^{\rm max})$, 
as a function of $E_e$. The solid line has $\Delta E=4.4$ eV as
per Ref.~\cite{weinplb}; 
the dashed line also includes recoil corrections as per Eq.~(\ref{recsub}). 
For reference the 
change in the theoretical form of the energy spectrum assumed
in Ref.~\cite{weinplb}, if $m_\nu^2 =0 \to - 4$ eV$^2$, 
is shown as the dot-dashed line as well. 
\label{fig:cfmnu}}
\end{center}
\end{figure}
\begin{figure}[ht]
\begin{center}
\includegraphics[width=4.6in]{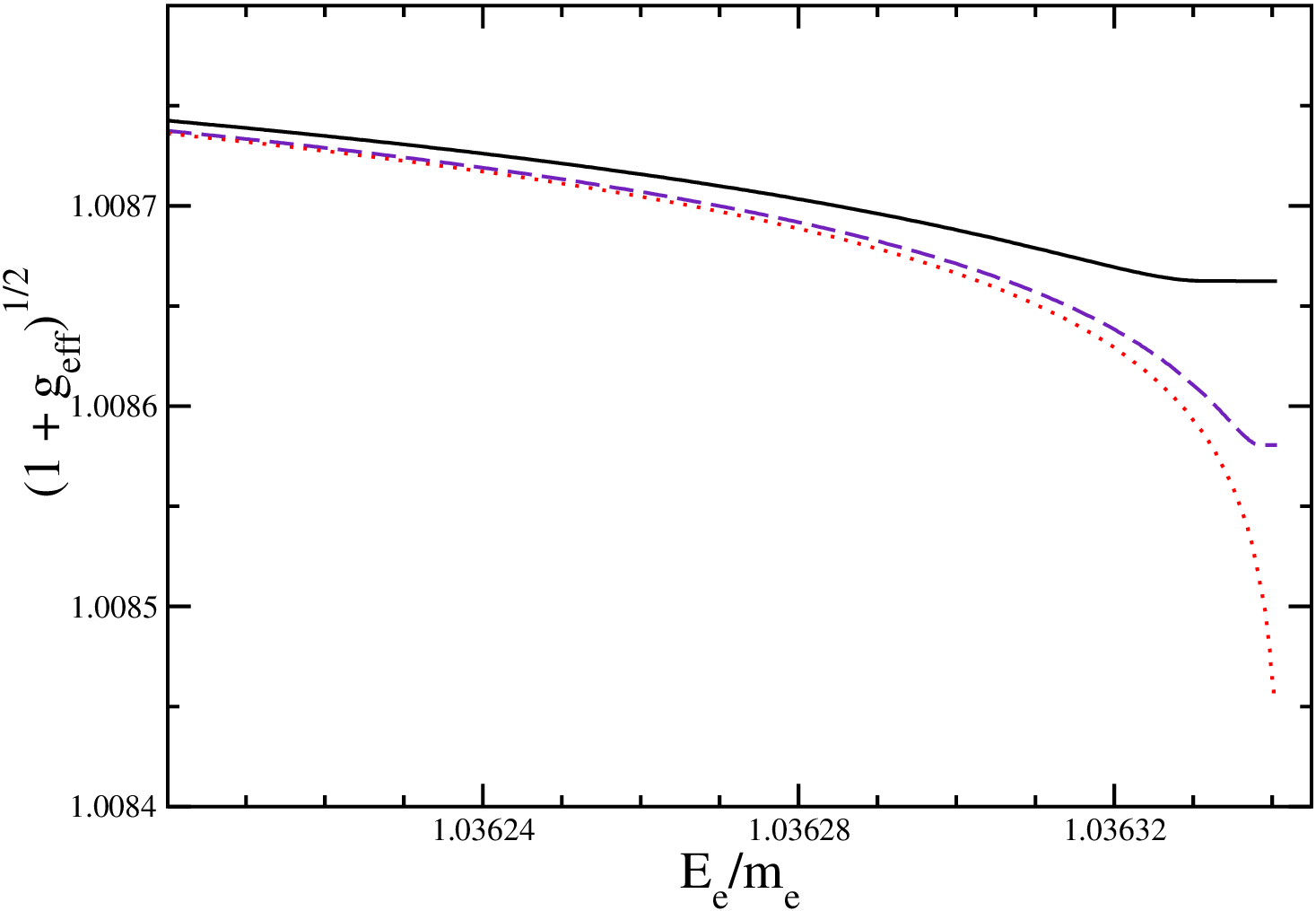}
\vspace{-0.2cm}
\caption{The ratio of the corrected to uncorrected Kurie plots,
namely $(1 + g_{\rm eff})^{1/2}$, with 
$g_{\rm eff} \equiv {\cal R} + (\alpha/2\pi)g(\Delta E, E_e, E_e^{\rm max})$, 
as a function of $E_e$. The solid line has $\Delta E=4.4$ eV, 
the dashed line has $\Delta E=1$ eV, and the dotted line has 
$\Delta E=0.1$ eV. 
\label{fig:kurie}}
\end{center}
\end{figure}

Armed with these results, we now proceed to evaluate 
the change in the electron energy spectrum upon the
inclusion of the outer radiative correction, 
$g(\Delta E, E_e, E_e^{\rm max})$. 
We illustrate 
$\Delta(d\Gamma/dE_e)=(d\Gamma_0/dE_e)\alpha g(\Delta E, E_e, E_e^{\rm max})
/(2\pi)$ in Fig.~\ref{fig:cfmnu}, using the energy resolution of the 
Mainz experiment~\cite{weinplb}, $\Delta E=4.4$ eV. In this figure, 
the inclusion of the $\Delta E$ dependence is of 
little impact; the resulting curve is hardly distinguishable 
from that which results from the use of Sirlin's 
function, Eq.~(\ref{sirlin}).  The recoil corrections are included
as well, so that we employ 
$d\Gamma/dE_e=d\Gamma_0/dE_e(1 + {\cal R} + F + 
\alpha/(2\pi) g(\Delta E, E_e,E_e^{\rm max}))$;
they are rather small, though they are appreciable. 
The analysis of Ref.~\cite{weinplb} assumes the theoretical
form given in Eq.~(\ref{fermiform}), inferring 
$m_\nu^2 = - 3.7 \pm 5.3 \pm 2.1$ eV$^2$ 
from their experimental data. Thus, for reference, 
we also show
$\Delta(d\Gamma/dE_e)=d\Gamma_{F} (m_\nu^2 = -4 \hbox{eV}^2)/dE_e
- d\Gamma_{F} (m_\nu^2 = 0 \hbox{eV}^2)/dE_e$
in Fig.~\ref{fig:cfmnu}. 
Note that employing 
$m_\nu^2 >0$ yields a $\Delta(d\Gamma/dE_e)$ which differs in 
sign from that generated with $g(\Delta E,E_e, E_e^{\rm max})$.
It is apparent that the change in the 
theoretical energy spectrum due to the neglected ${\cal O}(\alpha)$
correction acts to increase the electron energy spectrum; this effect
is also realized through a negative value of $m_\nu^2$ in 
Eq.~(\ref{fermiform}). The neglect of the outer radiative
correction generates a {\it negative} shift in $m_\nu^2$.    
We emphasize that this shift is a consequence of the change
in shape in the electron 
energy spectrum induced by the outer radiative correction. 
We illustrate this in Fig.~\ref{fig:kurie}, in which we show 
the ratio of the corrected to uncorrected Kurie plots, recalling
that the Kurie plot is 
$K(E_e)\equiv [(d\Gamma/dE_e)(1/Fp_eE_e)]^{1/2}$ versus $E_e$. 
This ratio is simply $(1 + g_{\rm eff})^{1/2}$ to ${\cal O}(\alpha)$, 
where 
$g_{\rm eff} \equiv {\cal R} + (\alpha/2\pi)g(\Delta E, E_e, E_e^{\rm max})$.

\begin{figure}[ht]
\begin{center}
\includegraphics[width=4.6in]{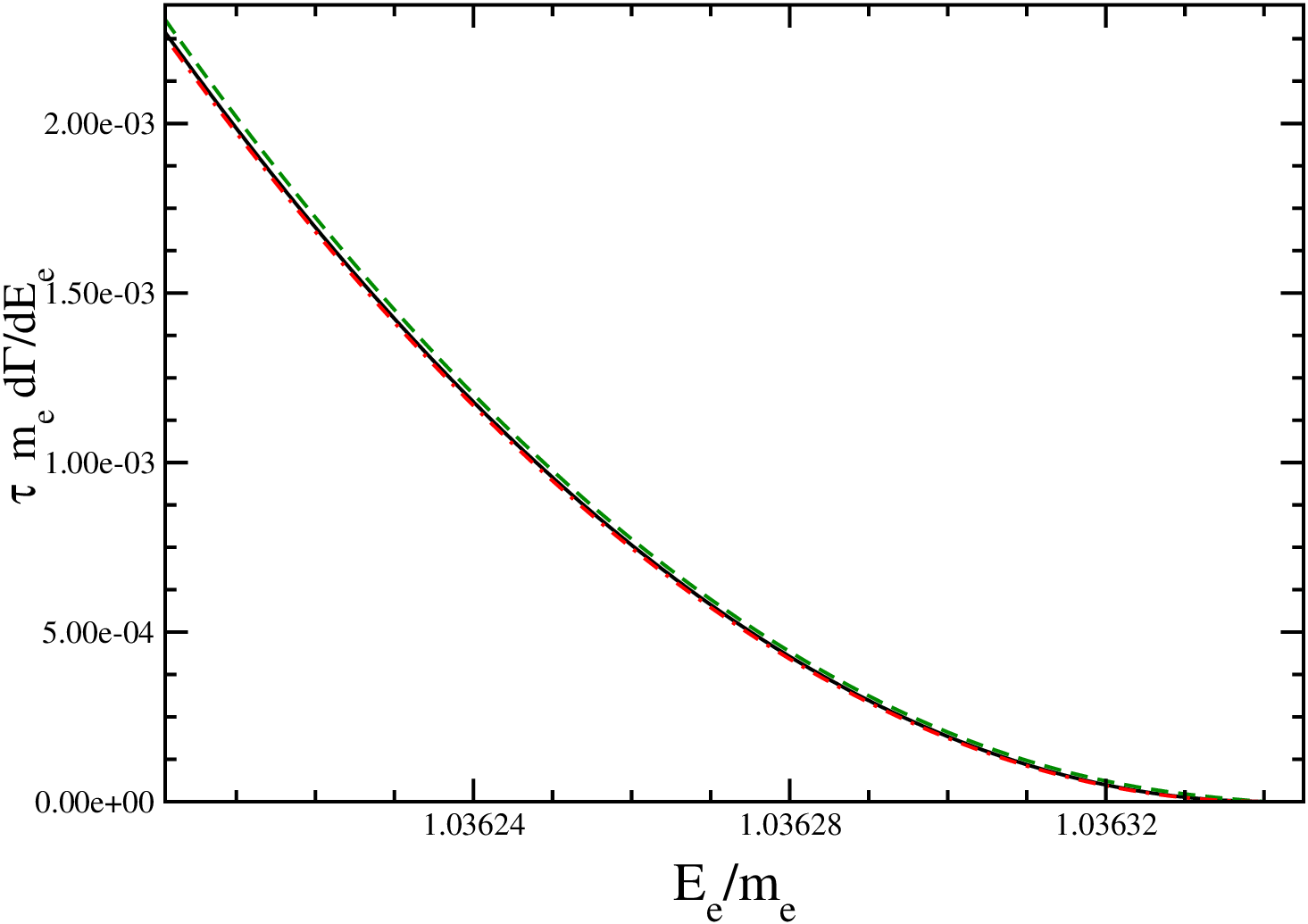}
\vspace{-0.2cm}
\caption{The electron energy spectrum for 
$E_e^{\rm max} - E_e \le 70$ eV, as a function of $E_e$. 
The solid curve shows the theoretical spectrum to be fit, which 
includes 
$g(\Delta E,E_e,E_E^{\rm max})$ with $\Delta E$=4.4 eV and
recoil corrections as per Eqs.~(\ref{recsub},\ref{recoil}).  
The dashed curve is realized from Eq.~(\ref{fit}) using 
$E_e^{\rm max}=18.57$ keV, $A=3.55020\cdot 10^5$, and 
$m_\nu^2 = -67$ eV$^2$. In constrast, the dot-dashed curve has 
$A=3.49446\cdot 10^5$ and $m_\nu^2=-0.01$ eV$^2$. 
\label{fig:fit}}
\end{center}
\end{figure}

To estimate the impact of the remaining ${\cal O}(\alpha)$ 
correction on the value of the neutrino mass, we introduce the fit function
used in earlier work~\cite{weinplb}: 
\begin{equation}
\tau m_e \frac{d\Gamma}{dE_e} = A m_e^{-4}
F(Z,R_{\rm e},E_e) 
p_e E_e (E_e^{\rm max} - E_e)\sqrt{(E_e^{\rm max} - E_e)^2 - m_\nu^2} + B\,,
\label{fit}
\end{equation}
where $A$, $E_e^{\rm max}$, $m_\nu^2$, and $B$ are all fit to
the electron energy spectrum. 
For our comparison, which we effect for purposes of illustration, 
we have neglected the contributions associated
with the excited 
final states of 
the daughter $^3$He$^+$ - T molecule and include
the elastic contribution only. 
Note that $B$ represents a constant
experimental background, so that we set $B=0$. 
Fitting the remaining 
parameters in Eq.~(\ref{fit}) to the dashed curve in Fig.~\ref{fig:cfmnu} 
yields the comparisons shown in Fig.~\ref{fig:fit}. We fit
the last 70 eV of the electron energy spectrum and set 
$\Delta E$=4.4 eV, in an attempt to simulate the 
conditions in Sec.~6.3 of Ref.~\cite{weinplb}, for which 
$m_\nu^2 = - 3.7 \pm 5.3 \pm 2.1$ eV$^2$ was inferred. 
We fit the quantity $\tau m_e d\Gamma/dE_e$, where the lifetime
$\tau$ is $\tau=t_{1/2}/\ln{2}$, noting, 
for reference, that $A=3.43537\cdot 10^5$ in our theoretical curves. 
Using the MIGRAD minimization program in the CERN package ROOT~\cite{root}, 
we find two different fits of comparable quality, which 
possess very different values of $m_\nu^2$; apparently a significant 
shift in $m_\nu^2$ can be accommodated by 
normalizations $A$ which differ by $\sim 2$\%. 
In both cases $m_\nu^2 <0$; we cannot successfully fit our curves using
a non-negative value of $m_\nu^2$ in Eq.~(\ref{fit}). In constrast, if we
fit Eq.~(\ref{fit}) to a curve containing the Fermi function only, 
these features do not occur. In this case, 
we find fits with $|m_\nu^2| < 1$ eV$^2$ are consistent with the input curve;
$m_\nu^2$ need not be negative definite. 

The refinements we have introduced, namely, the $\Delta E$ dependence
to the outer radiative corrections cum the recoil corrections, 
shift $m_\nu^2$ at no larger than the ${\cal O}(1\, \hbox{eV}^2)$ level. 
As Fig.~\ref{fig:kurie} makes clear, 
this conclusion is sensitive to the precise value of $\Delta E$, as
well as the interval in $E_e$ over which the neutrino mass is fit. 
For other choices of these parameters, their relative impact could be more
significant. Given the results shown in Fig.~(\ref{fig:fit}), 
we cannot make a robust conclusion concerning the absolute
scale of the shift in $m_\nu^2$ our previously neglected 
theoretical corrections would induce in a more realistic analysis; 
nevertheless, we can say that the consequence 
of neglecting these terms is to push $m_\nu^2<0$ in an artificial way. 
We presume that with the replacement of Eq.~(\ref{fit}) with a fit
function incorporating the radiative and recoil 
corrections we have calculated such artificial shifts would disappear. 
In the analysis we have effected, this turns out to be the case. 

In this letter we have evaluated the ${\cal O}(\alpha)$ outer 
radiative correction, $g(\Delta E,E_e,E_e^{\rm max})$, 
to the electron energy spectrum in $^3$H $\beta$-decay, so 
that $(\alpha/2\pi)g + F -1$, where $F$ is the Fermi function, constitutes
the complete ${\cal O}(\alpha)$ correction to the electron
energy spectrum. We have updated the calculation of Sirlin~\cite{sirlin}
to include the dependence of the outer radiative correction
on the detector energy resolution $\Delta E$; as a consequence
the ${\cal O}(\alpha)$ correction 
we compute to the shape of the electron energy
spectrum is {\it finite} as $E_e \to E_e^{\rm max}$. 
However, as necessary, 
it has no impact on the radiative correction to 
the total decay rate. Interestingly, the outer radiative correction 
was omitted all together in earlier studies of tritium 
$\beta$-decay~\cite{robprl,exp1,exp2,exp3,wein94,exp4,troitsk,weinplb,katrin}; 
we have shown that the shape correction associated with 
this $\sim 2\%$ shift mimicks a negative value of $m_\nu^2$. 
We believe it is necessary to update earlier
experimental analyses to take this theoretical correction into
account, to realize an accurate determination of the neutrino mass. 
A highly accurate theoretical spectrum 
can be found by modifying Eq.~(\ref{fermiform}), the form
used in earlier experimental analyses of $^3$H $\beta$-decay,
through the substitution $F \to F^\ast + 
(\alpha/2\pi)g(\Delta E,E_e,E_e^{\rm max})
+ {\cal R}$, using Eqs.~(\ref{ints}-\ref{res3}) and Eq.~(\ref{recoil}).
Our focus has been on $g(\Delta E,E_e,E_e^{\rm max})$ and 
${\cal R}$; theoretical corrections to these terms 
accrue from 
i) ${\cal O}(Z\alpha^2)$ corrections, which are known~\cite{sirzuc}, 
and ii) ${\cal O}(1\%)$ corrections to the recoil-order term, 
Eq.~(\ref{recoil}), but such corrections would appear beyond
the scope of current and planned experiments.
The corrected Fermi function $F^\ast$, which includes 
corrections such as those due to the finite nuclear size 
and to charge screening of the nuclear charge by atomic electrons, 
is 
detailed in Ref.~\cite{Wilkinson}; recoil corrections~\cite{recoil} and
outer radiative corrections, as calculated by Sirlin~\cite{sirlin}, 
are considered in this reference as well. The outer radiative
corrections are the largest of these corrections~\cite{Wilkinson}. 
Realistic experimental
conditions demand that Eq.~(\ref{fit}), as well as the fit form
we advocate, be adapted to include the population of all the excited 
final states $i$ of the daughter He$^+$-T molecule; the
excitation energies and amplitudes are computed from atomic
theory. As a consequence, the elucidation of the neutrino
mass in $^3$H $\beta$-decay relies on atomic physics input
which cannot be wholly subjected to exhaustive, independent empirical test. 
Nevertheless, from the viewpoint of the theoretical radiative and recoil
corrections,  a sub-eV determination of the neutrino
mass should be possible.

\smallskip
\noindent{\large {\bf Acknowledgements}}

\smallskip\noindent
The work of S.G. is supported in
part by the U.S. Department of Energy under contract number 
DE-FG02-96ER40989. We thank Wolfgang Korsch for helpful discussions
and much-needed assistance with ROOT and Christian
Weinheimer for useful comments and correspondence. S.G. thanks
the SLAC theory group for hospitality during the completion
of this manuscript. We are grateful to Burton Richter for 
suggestions
helpful in improving our presentation. 


\end{document}